\documentclass[fleqn,usenatbib]{mnras}

\usepackage{newtxtext,newtxmath}
\usepackage{times}

\usepackage{graphicx}	
\usepackage{amsmath}	
\usepackage{amssymb}	
\usepackage{comment}

\title[Constraints on DM from flux-ratio anomalies]{SHARP -- VII. New constraints on the dark matter free-streaming properties and substructure abundance from gravitationally lensed quasars}

\author[J.-W. Hsueh et al.]{J.-W. Hsueh,$^{1,2}$\thanks{E-mail: hsueh@astro.rug.nl} 
W. Enzi,$^{3}$ 
S. Vegetti,$^{3}$ 
M. W. Auger,$^{4}$ 
C. D. Fassnacht,$^{2}$ 
G. Despali,$^{3}$
\newauthor 
L. V. E. Koopmans$^{1}$ 
and J. P. McKean$^{1,5}$
\\
$^{1}$Kapteyn Astronomical Institute, University of Groningen, P.O.Box 800, 9700AV, Groningen, the Netherlands\\
$^{2}$Department of Physics, University of California, Davis, USA\\
$^{3}$Max Planck Institute for Astrophysics, Karl-Schwarzschild-Strasse 1, D-85740 Garching, Germany\\
$^{4}$Institute of Astronomy, University of Cambridge, Madingley Road, Cambridge, CB3 0HA, UK\\
$^{5}$ASTRON, Netherlands Institute for Radio Astronomy, P.O. Box 2, 7990 AA Dwingeloo, the Netherlands\\
}

\date{Accepted XXX. Received YYY; in original form ZZZ}

\pubyear{2019}

\begin{document}
\label{firstpage}
\pagerange{\pageref{firstpage}--\pageref{lastpage}}
\maketitle

\begin{abstract}
We present an analysis of seven strongly gravitationally lensed quasars and the corresponding constraints on the properties of dark matter. Our results are derived by modelling the lensed image positions and flux-ratios using a combination of smooth macro models and a population of low-mass haloes within the mass range $10^6$ to $10^9$~M$_\odot$. Our lens models explicitly include higher-order complexity in the form of stellar discs and luminous satellites, as well as low-mass haloes located along the observed lines of sight for the first time. Assuming a Cold Dark Matter (CDM) cosmology, we infer an average total mass fraction in substructure of $f_{\rm sub} = 0.012^{+0.007}_{-0.004}$ (68~per cent confidence limits), which is in agreement with the predictions from CDM hydrodynamical simulations to within 1$\sigma$. This result is closer to the predictions than those from previous studies that did not include line-of-sight haloes. Under the assumption of a thermal relic dark matter model, we derive a lower limit on the particle relic mass of $m_{\rm th} > 5.58$ keV (95 per cent confidence limits), which is consistent with a value of $m_{\rm th} > 5.3$ keV from the recent analysis of the Ly$\alpha$ forest.  
We also identify two main sources of possible systematic errors and conclude that deeper investigations in the complex structure of lens galaxies as well as the size of the background sources should be a priority for this field.
\end{abstract}

\begin{keywords}
lensing: strong -- cosmology: dark matter -- galaxies: structure
\end{keywords}


\section{Introduction}

Strong gravitational lensing has been shown to be a powerful tool to probe the presence of low-mass haloes in distant galactic haloes \citep{Dalal2002,V09,V10a,V10,V12,N14,Hezaveh16,Gilman2018,Bayer2018}, and along their lines of sight \citep{Metcalf05,Despali2018,Gilman2019}. {These low-mass haloes are dark matter dominated and therefore free from the uncertainty of baryonic process during structure formation.} Hence, comparing to other approaches that focus on the local Universe, strong gravitational lensing provides an independent and promising approach to differentiate between alternative dark matter theories that modify the linear matter power-spectrum and result in a different amount of low-mass haloes, such as cold \citep[CDM,][]{Springel2008}, warm \citep[WDM,][]{Sch2012,Lovell2014} or fuzzy dark matter \citep{Hui2017,Robles2019}. Mainly two approaches have been used to detect these low-mass haloes using strong gravitational lensing observations, which we now review. 

The {\it gravitational imaging technique} constrains the projected position and effective mass of individual low-mass haloes via their effect on the surface brightness distribution of extended arcs \citep{K05,V09}. This technique has, so far, led to the detection of  a few haloes in the mass regime between $10^8$ and $10^9$~M$_\odot$ with optical/infrared (IR) \citep{V10a,V10,V12} and sub-millimetre imaging \citep{Hezaveh16}, while future observations with long baseline interferometers are expected to lead to the discovery of haloes with masses lower than $10^7$~M$_\odot$ \citep{SKA}. Recently, \citet{V14a,V18} and \citet{Ritondale2018} have used samples of 10--20 lenses from the SLOAN ACS Lens Survey \citep[SLACS;][]{Bolton2006} and the  BOSS Emission Line Lens Survey \citep[BELLS;][]{Shu2016} to constrain the halo mass function in the regime between $10^8$ and $10^{10}$~M$_\odot$. They found that their results are consistent with the predictions from the CDM paradigm \citep[e.g. ][]{Xu15,Despali2017}, but more observations of higher quality are required to potentially rule out alternative WDM models.

The analysis of gravitationally lensed quasars uses the flux-ratio relation between the merging images to probe the total amount of low-mass haloes without inferring their  individual positions and masses. 
Specifically, the presence of low-mass haloes is expected to change the relative fluxes of the multiple images compared to predictions from smooth lensing potentials. Currently, only those gravitational lens systems that produce four images of the background quasar can be used, as they provide enough constraints  on the lens macro-model (typically, a smooth mass model plus an external shear). \citet{Mao1998} and \citet{metcalf01} were the first to suggest that flux-ratio anomalies could be related to the presence of dark substructures contained within the dark matter halo of the foreground lensing galaxy, and therefore, these systems could be used to constrain the substructure fraction of distant galaxies. Follow-up studies, based on observations and numerical simulations, corroborated this idea and explicitly demonstrated the feasibility of flux-ratio anomalies as a means to detect low-mass substructure \citep{Bradac02,Metcalf02,Dobler2006,N14}. 

\citet{Dalal2002} presented the first homogeneous analysis of a small sample of seven lensed quasars, obtaining a result that was marginally consistent with the CDM paradigm. However, this result was contested first by \citet{Mao2004} and later by \citet{Xu09,Xu15}, who showed that the level of the observed flux-ratio anomalies was significantly higher than expected from the subhalo population in high-resolution numerical simulations. Instead, they suggested that either line-of-sight haloes or complex mass distributions of the lensing galaxies were more likely the cause of the observed signal \citep[see also][]{Moller03,Quadri2003,Chen2003,Metcalf05,Inoue2012}. Recently, significant progress has been made towards a better understanding of the origins of flux-ratio anomalies in these systems. In particular, \citet{Hsueh2016,Hsueh2017,Hsueh2018} and \citet{Gilman2018} have shown that baryonic structures, such as stellar discs, are a likely source of extreme flux-ratio anomalies, and it was demonstrated that deep-imaging observations were needed to break the degeneracy between low-mass haloes and other complexity in the lens mass distribution. Using mock observations, \citet{Gilman2019} have shown the contribution from line-of-sight haloes to be significant and to provide extra constraining power on the halo mass function and the properties of dark matter \citep[see also][]{Metcalf05,Despali2018}.

Since the properties of dark matter are inferred from deviations between the observed flux ratios and those predicted by the gravitational lensing mass model, reliable measurements of the flux ratios are needed. Historically, the analysis of gravitationally lensed quasars was restricted to systems with radio and mid-infrared (MIR) observations. This is because the radio jets produced from synchrotron emission and the thermal emission from the dusty torus of lensed quasars are expected to be free from dust extinction and stellar micro lensing \citep[however, see][for an exception]{Koopmans2000}.  The small number of radio-loud lensed quasars and the difficulty in obtaining high-resolution MIR imaging from ground-based telescopes have limited the size of suitable samples. However, recent studies have shown a possible way forward to increase the sample size in the short-term. In particular, \citet{N14} have demonstrated that narrow emission lines provide a new avenue for flux-ratio studies with near infrared spectroscopy \citep[see also][]{MoustakasMetcalf2003}, while \citet{Stacay2018} have presented a new approach based on observations of cold dust and CO emission lines with the Atacama Large Millimetre/submillimetre Array (ALMA; see also \citealt{Inoue17}). The launch of the {\it James Webb Space Telescope} ({\it JWST}) will also make the MIR flux measurements of lensed quasars faster and easier to obtain \citep{JWST}, while future large-scale surveys, such as with the Large Synoptic Survey Telescope \citep[LSST;][]{LSST}, {\it Euclid} \citep{Cimatti2012}, and the Square Kilometre Array \citep[SKA;][]{SKA} are expected to lead to the discovery of thousands of new gravitationally lensed quasars.

In this paper, we present a new analysis of the current sample of four-image gravitationally lensed quasars that have well-studied lens models and reliable radio or MIR flux measurements. For the first time, our analysis includes not only the contribution of substructure within the lensing galaxy, but also that of stellar discs and line-of-sight haloes. Moreover, we use improved measurements of the observed flux-ratios that have been obtained from monitoring campaigns to derive tighter constraints on the substructure and halo mass functions, and thereby the free-streaming properties of dark matter. In particular, our analysis focuses on thermally produced dark matter of which weakly interacting particles (WIMPs) are the best theoretically motivated CDM candidate.
Our paper is organized as follows. In Section \ref{sec:Methodology}, we introduce our Bayesian modelling technique, while in Section \ref{sec:data}, we describe the observational data used in the analysis. In Section \ref{sec:results}, we present our new constraints on the dark matter mass function and discuss the implications for dark matter physics. In the first part of our analysis, we present the impact of line-of-sight haloes on the substructure mass function inference, assuming a concordant $\Lambda$CDM cosmology. The second part of our analysis provides constraints on the mass of a thermal relic dark matter particle. In Section \ref{sec:conclusions}, we summarize our results and discuss future extensions to this work.

Throughout out, we assume a flat cosmology with $\Omega_{\rm M} = 0.28$ and $H_0 = 70$~km\,s$^{-1}$~Mpc$^{-1}$.


\section{Methodology}
\label{sec:Methodology}
In this section, we describe the Bayesian methodology used to infer the low-mass-end of the halo mass function and the underlying dark matter properties. Specifically, we introduce the properties of the substructure and the line-of-sight halo populations in Sections \ref{sec:Substructure} and \ref{sec:LOS}, respectively. We discuss the specifics of the macro models in Section \ref{sec:macro}. In Section \ref{sec:Bayesian_inference}, we present the posterior probability of the dark matter parameters, given the observed data, and in Section \ref{sec:MCMC} we describe our analysis strategy.

\begin{figure}
\centering
\includegraphics[scale=0.5]{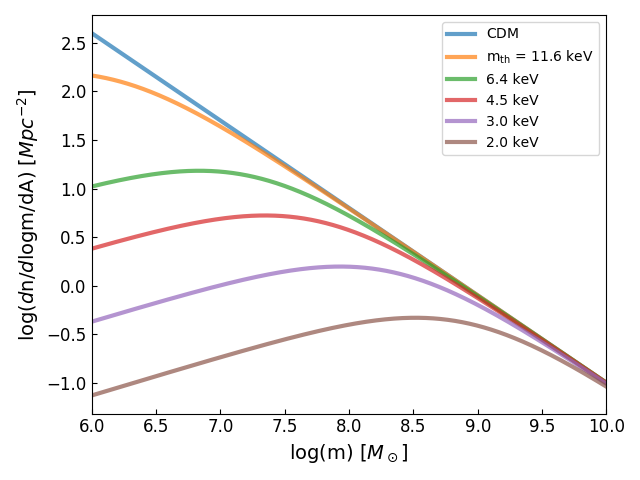}
\caption{The subhalo mass functions of CDM and WDM models with different thermal relic particle mass, $m_{\rm th}$. WDM models with a larger free streaming length, that is, smaller particle mass, lead to a suppression in the number of substructures at progressively larger substructure masses.}
\label{fig:mf}
\end{figure}

\subsection{The substructure population}
\label{sec:Substructure}

We assume the substructure population to be well described by spherical NFW haloes \citep*{Navarro1997} with virial masses between $10^6$ and $10^9$~M$_\odot$, and a concentration-mass relation from \citet{Duffy2008}. We neglect the effect of tidal truncation and changes in the concentration-mass relation as a function of distance from the host halo centre, as both effects have been shown to be of secondary importance in terms of the lensing effects of these low-mass haloes \citep{Despali2018}. Following \citet{Sch2012} and \citet{Lovell2014}, we parameterize the substructure mass function (i.e., the number density of substructures in the mass range $m,m+dm$ per unit area) as,
\begin{equation}
n_{\rm sub} (m) = \frac{d^2 N_{\rm sub} (m)}{dm~dA} = n_{\rm 0}
\left(\frac{m}{m_0}\right)^{-1.9}\left(1+\frac{M_{\rm{hm}}}{m}\right)^{~-1.3}\,,
\label{equ:sub_dn}
\end{equation}
where $M_{\rm{hm}}$ is the half-mode mass, that is, the mass scale at which the transfer function is suppressed by 50 per cent relative to CDM \citep[$M_{\rm{hm}} = 0$ for idealized CDM and $M_{\rm{hm}} \sim 10^{-6}M_\odot$ for WIMPs,][]{Schneider2013}.  The normalization constants $n_0$ and $m_0$ can be related to the projected total mass fraction in subhaloes defined below.
For thermal relic dark matter models, the half-mode mass is related to the mass of the dark matter particle by,
\begin{equation}
M_{\rm{hm}} = \frac{4\pi}{3}\bar{\rho}\Big( 6.97 \lambda^{\rm{eff}}_{\rm{fs}} \Big)^3
\label{equ:mhm_ms}
\end{equation}
\citep{Viel2005,Sch2012}, where $\bar{\rho} = \Omega_M \cdot \rho_{\rm crit}$ is the background density of the universe and $\lambda^{\rm{eff}}_{\rm{fs}}$ is the effective free-streaming length scale, which is given by,
\begin{equation}
    \lambda^{\rm{eff}}_{\rm{fs}} = 0.049~ \Big( \frac{m_{\rm{th}}}{keV} \Big)^{-1.11} \Big( \frac{\Omega_{\rm{th}}}{0.25} \Big)^{0.11} \Big(\frac{h}{0.7} \Big)^{1.22} ~{\rm Mpc~h^{-1}},
\end{equation}
where $m_{\rm th}$ is the thermal relic mass of the dark matter particle.\\
The inverse conversion between $m_{\rm th}$ and $M_{\rm hm}$ can be expressed as \citep{Nadler2019}:
\begin{equation}
    m_{\rm th} = 2.32~ \Big( \frac{M_{\rm hm}}{10^9 M_\odot} \Big)^{-0.3} \Big( \frac{\Omega_{\rm th}}{0.25}  \Big)^{0.4}\Big(\frac{h}{0.7} \Big)^{0.8} ~{\rm keV}.
\end{equation}
In Figure \ref{fig:mf}, we show the substructure mass function for the idealized CDM and thermal relic models of different particle mass. 

As lensing is sensitive to the total projected mass distribution within a cylinder, we define $f_{\rm sub}$ as the ratio between the total mass in substructure within a projected cylinder with a radius twice as large as the main lens Einstein radius, $\theta_E$, and the total mass of the main halo within the same projected cylinder.  That is,
\begin{equation}
f_{\rm sub} = \frac{ \Sigma m_{\rm sub,proj}}{M_{\rm proj}} = \frac{ \int ^{M_{\rm{high}}}_{M_{\rm{low}}} {n_{\rm sub} (m)}~m ~dm \cdot A_{\rm proj}}{M_{\rm proj}},
\label{eq:fsub}
\end{equation}
where $A_{\rm proj}$ is the area within the aperture of the projected cylinder and $(M_{\rm{low}},M_{\rm{high}}) = (10^6,10^9)~$M$_\odot$. The expectation value of substructures, $\mu_{\rm sub}$, within the projected cylinder is expressed as,
\begin{equation}
\mu_{\rm sub} =  \int^{M_{\rm{high}}}_{M_{\rm{low}}} {n_{\rm sub} (m)} ~dm \cdot A_{\rm proj}.
\label{equ:norm}
\end{equation}
Following \citet{Xu15}, we assume the projected position of substructure on the plane of the lensed images to be uniform within $2\theta_E$.

\citet{Gao2004} and \citet{Xu15} have shown the substructure mass fraction to be a function of the host halo virial mass. Because we do not have virial masses for the lensing galaxies in our sample, we neglect this dependence. As a consequence, our constraints on $f_{\rm sub}$ should be interpreted as a mean value. While our sample is certainly not homogeneous in this respect, it is also unlikely to span a wide range of virial masses, as, statistically speaking, strong gravitational lens galaxies are more likely to reside in haloes of about $10^{13}$~M$_\odot$ \citep[e.g.,][]{Sonnenfeld2018}. It should be also noted that we ignore the redshift dependence on $f_{\rm sub}$ among our sample since the systematic errors are considered to be larger than the effect of redshift evolution.

\subsection{The line-of-sight halo population}
\label{sec:LOS}

Similar to what has been done for the substructure halo population, we include line-of-sight haloes as spherical NFW haloes with virial masses between $10^6$ and $10^9$~M$_\odot$, and again use the concentration-mass relation from \citet{Duffy2008}. We apply the \citet{Sheth1999} halo mass function to calculate the number density of haloes per co-moving volume and within the mass range $m, m+dm$,  and the best-fitting parameters are optimized for the Planck cosmology \citep{Despali2016}. We include the effect of the free-streaming properties of the dark matter particles with the same factor used for the substructure mass function, that is, an attentuation with $\left(1+M_{\rm hm}\right/m)^{-1.3}$. We assume the normalization of the halo mass function to be constant, that is, we assume an average number density of haloes and neglect fluctuations amongst the different lines of sight. We discuss the implications of this choice in Section \ref{sec:results}.

\begin{figure}
    \centering
    \includegraphics[scale=0.55]{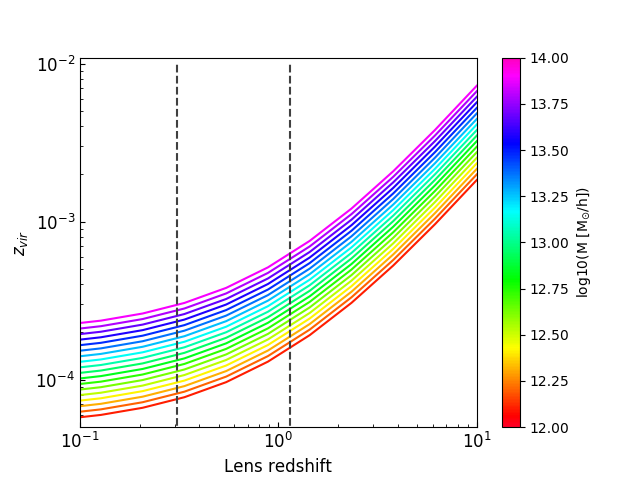}
    \caption{The evolution of the virial radius in terms of redshift for different values of the halo virial mass (in units of redshift). The two vertical lines indicate the lowest and highest lens redshift in our sample.}
    \label{fig:virz}
\end{figure}

We only consider line-of-sight haloes inside two light cones that share their base on the lens plane and have tips at the observer and at the redshift of the lensed quasar. The base of these cones is given by a circle of two times the Einstein radius, and is centred on the optical axis. To increase the computing efficiency, we consider multiple lens planes along the line of sight with an interval of $dz=0.05$, and on each redshift plane, the haloes have projected positions drawn from a uniform prior. Unlike substructures, we assume line-of-sight haloes to be located outside the virial radius of the lensing galaxy, that is, $z_{\rm los}>z_{l}+z_{\rm vir}$ or $z_{\rm los}<z_{l}-z_{\rm vir}$, with $z_{\rm vir}=10^{-3}$  (see Fig. \ref{fig:virz} for theoretical results and \citealt{Sonnenfeld2018} for more details about the relation between virial radius and stellar mass of the host galaxy).

\begin{table*}
\caption{Summary of the different parameters that enter our analysis.}
    \centering
    \begin{tabular}{c|c|c}
    \hline
    Parameter & Definition & Details\\
    \hline
     $\vec{\theta}$    & mass function parameters  & $f_{\rm sub}, M_{\rm hm}$; \\
     & & \\
     $f_{\rm sub}$ & mass fraction in substructure & $0.002 < f_{\rm sub} < 0.04$\\
     & w/i $2\theta_e$ projected area &\\
     & &\\
     $M_{\rm hm}$ & half-mode mass & $ 10^{5} < M_{\rm hm} < 10^9$~M$_{\odot}$\\
     & &\\
     $\vec{\theta_{\rm M}}$   & macro-model parameters  & SIE+$\Gamma$, source position, \\
     & & mass of luminous satellite or stellar disc\\
    && \\
    $\vec{\theta_{\rm m}}$ & {micro-model parameters} & substructure:  mass, position\\
    & & line-of-sight halo:  redshift, mass, position\\
    & &\\
    $\mathbf{d}$ & observed data & lensed image position, flux ratio\\
    & &\\
    $N$ & number of low-mass perturbers & $ N = N_{\rm sub} + N_{\rm los}$\\
    \hline
    \end{tabular}
    \label{tab:not}
\end{table*}

\subsection{The macro model}
\label{sec:macro}

Each lensing galaxy is modelled as a singular isothermal ellipsoid (SIE), with the contribution of an external shear component $\Gamma$. Note that higher order terms are also introduced to the macro model when a luminous satellite and/or a stellar disc is detected. For systems with a luminous satellite, we fix the centroid position of the satellite from optical/IR observations and assume a singular isothermal sphere (SIS) mass density profile. The Einstein radius of the luminous satellite is then the only free parameter. Similarly, for systems with a detected stellar disc, we introduce an exponential disc component and assign the mass of the disc as a free parameter, while the other parameters are kept fixed at the values inferred from the corresponding light distribution.

\subsection{Bayesian inference on dark matter}
\label{sec:Bayesian_inference}

In the following, we refer to the observed flux ratios, $f_i$, and positions, $\mathbf{x}$, of the lensed images, along with their uncertainties as the data, $\mathbf{d}$. The model parameters that we want to infer are the substructure mass fraction, $f_{\rm sub}$, and the half-mode mass, $M_{\rm hm}$; these are collectively referred to as the target parameters,  $\vec{\theta}$. We consider the macro-model parameters, that is, the lensing galaxy mass distribution and the source position,  $\vec{\theta}_{\rm M}$, as nuisance parameters. Further nuisance parameters are the number of substructures and line-of-sight haloes, $N$, and the micro-model parameters, $\vec{\theta}_{\rm m}$, which include their masses, projected positions and redshifts. Table \ref{tab:not} summarizes the notations and definitions used in this work.

Using Bayes theorem, we relate the posterior probability density of $\vec{\theta}$, marginalized over the nuisance parameters, to the likelihood function $P(\mathbf{d}|\vec{\theta},\vec{\theta}_{\rm M},\vec{\theta}_{\rm m},N)$ as,
\begin{multline}
P(\vec{\theta}|\mathbf{d}) \propto \sum_i^N \int  P(\mathbf{d}|\vec{\theta},\vec{\theta}_{\rm M},\vec{\theta}_{\rm m},N) P(\vec{\theta}_{\rm m}|\vec{\theta},\vec{\theta}_{\rm M}) P(N|\vec{\theta},\vec{\theta_{\rm M}})\\
\times P(\vec{\theta}_{\rm M})~P(\vec{\theta}) ~d\vec{\theta}_{\rm M}~d\vec{\theta}_{\rm m}\,.
\label{equ:posterior}
\end{multline}
Under the assumption of Gaussian errors on the observed fluxes and positions, the log-likelihood function is related to $\chi_{\rm tot}^2=\chi_{\rm flux}^2+\chi_{\rm pos}^2$, such that,
\begin{equation}
    P(\mathbf{d}|\vec{\theta},\vec{\theta}_{\rm M},\vec{\theta}_{\rm m},N) \approx \mathcal{G}(\chi_{\rm tot}^2(\vec{\theta},\vec{\theta}_{\rm M},\vec{\theta}_{\rm m},N)),
\end{equation}
where 
\begin{equation}
\mathcal{G} \propto \exp \left(-\frac{1}{2} \left(  \chi_{\rm flux}^2 +\chi_{\rm pos}^2 \right) \right).
\end{equation}
For $P(N|\vec{\theta},\vec{\theta}_{\rm M})$, we adopt a Poisson probability function with an expectation number of substructures, $\mu_{\rm sub}$, and line-of-sight haloes, $\mu_{\rm los}$, respectively. The contribution from each population is given by the integral over the respective mass functions as defined in Sections \ref{sec:Substructure} and \ref{sec:LOS}. We apply a Monte-Carlo approach to compute the integral in eq. (\ref{equ:posterior}). To increase the computing efficiency, we introduce importance sampling of the macro-model prior, $P(\vec{\theta}_{\rm M})$, as,
\begin{equation}
    P(\vec{\theta}_{\rm M}) \longrightarrow \frac{P(\vec{\theta}_{\rm M})}{Q(\vec{\theta}_{\rm M})} ~ Q(\vec{\theta}_{\rm M}),
\end{equation}
where $Q(\vec{\theta}_{\rm M})$ is obtained from an MCMC modelling of the data under the assumption of $N=0$. We then draw $\vec{\theta}_{\rm M}$ realisations from $Q(\vec{\theta}_{\rm M})$. The likelihood of each realisation is then weighted by $P(\vec{\theta}_{\rm M})/Q(\vec{\theta}_{\rm M})$. For the micro-model parameters, we adopt priors as discussed in Sections \ref{sec:Substructure} and \ref{sec:LOS}, respectively. 

In the following subsection, we provide more details on the ray-tracing strategy that we have adopted to compute the likelihood and the posterior probability functions.

\begin{figure}
    \centering
    \includegraphics[scale=0.4]{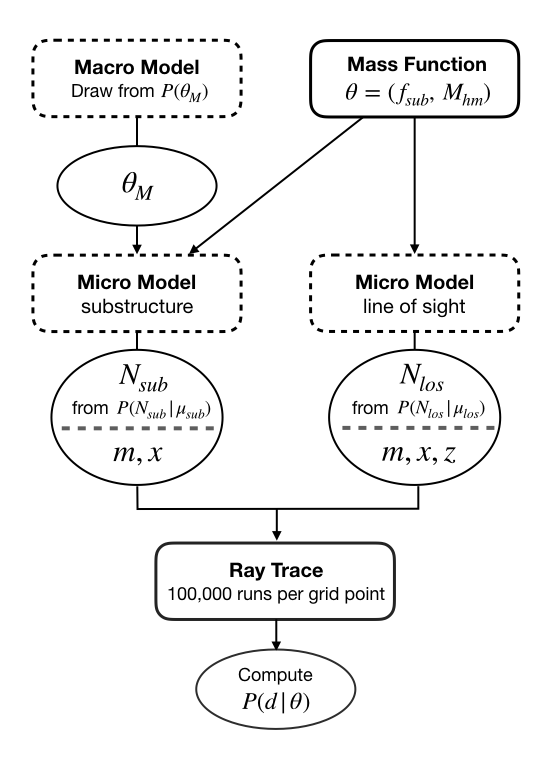}
    \caption{A schematic view of our analysis strategy. On each grid point of target parameters $(f_{\rm sub},M_{\rm{hm}})$ we run 100\,000 simulations to collect statistical inference by comparing the predicted lens observable with the measured ones.}
    \label{fig:flow}
\end{figure}

\subsection{Analysis scheme}
\label{sec:MCMC}

\begin{figure}
    \centering
    \includegraphics[scale=0.5]{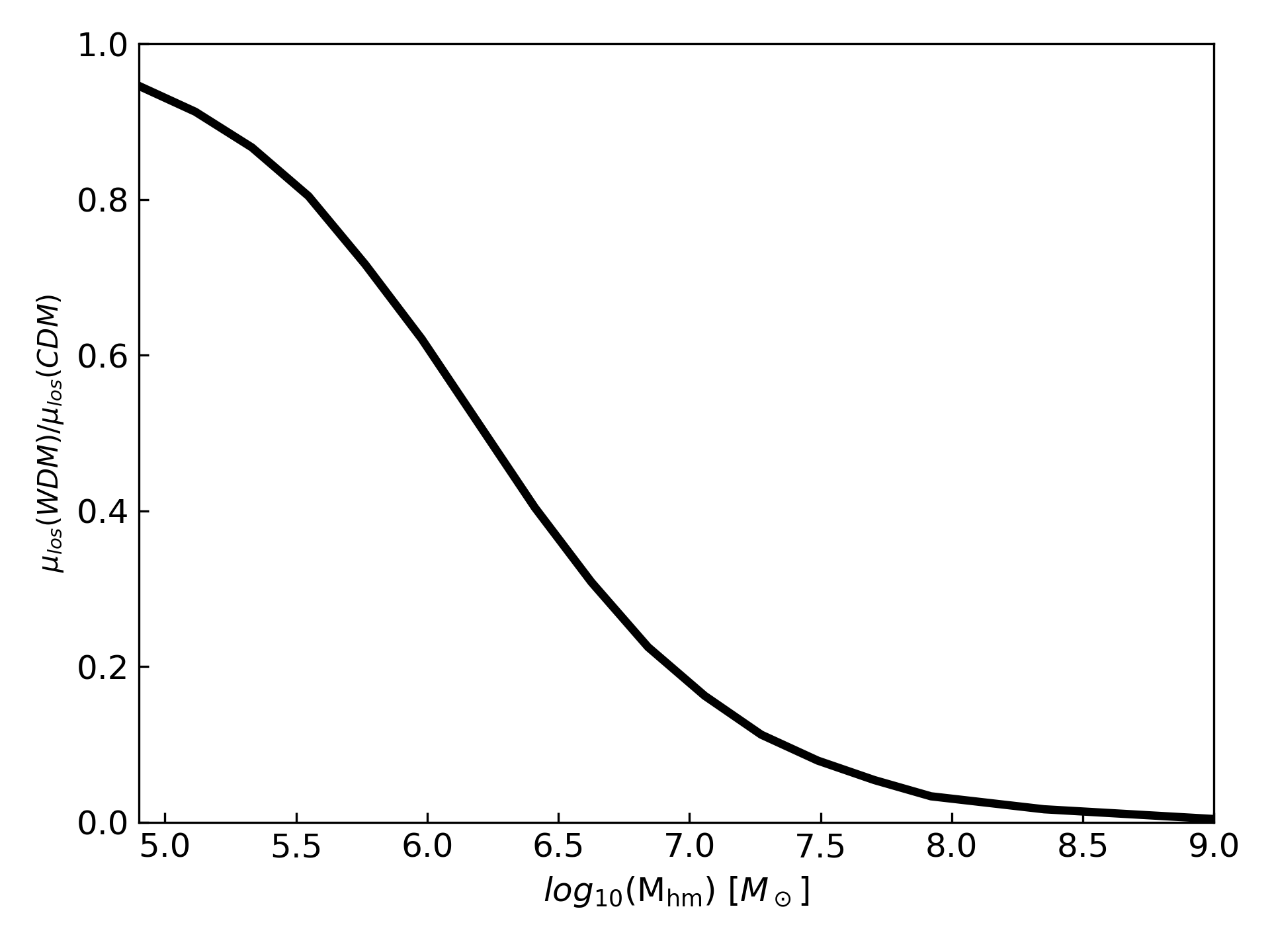}
    \caption{The ratio between the expected number of detectable line-of-sight haloes in WDM models of different half-mode mass $M_{\rm hm}$ and idealized CDM. The expected number of perturbers are obtained by integrating equation~(\ref{equ:norm}) between $10^6 - 10^9~ M_{\odot}$, i.e. the range of halo mass that can be detected with the considered data.}
    \label{fig:muratio}
\end{figure}

 We assume a uniform prior in $\log$ space for the target parameters in the ranges $0.2~{\rm per~cent}<f_{\rm sub}<4~{\rm per~cent}$ and $10^{5}<M_{\rm hm}<10^9$~M$_{\odot}$. We achieve this by defining a regular grid in logarithmic space where all the points are equally weighted. The lower bound of the $M_{\rm hm}$ prior is due to the fact that the likelihood function flattens to a constant value below $\sim10^{5}~\textrm{M}_\odot$, as the data lose sensitivity to the small difference in the number of detectable perturbers predicted by the different dark matter models (see Figure~\ref{fig:muratio}). Note that we also evaluate the CDM WIMP model at $M_{\rm hm}=10^{-6}$~M$_{\odot}$ and interpolate the likelihood for values up to $M_{\rm hm}=10^{5}$~M$_{\odot}$. The loss of sensitivity is related to the size and structure of the background source and the opening angle of the merging images (we refer to Section \ref{sssec:source} for a more detailed discussion). From Figure~\ref{fig:muratio}, it can bee seen that for dark matter models with $M_{\rm hm} > 10^8~M_{\odot}$ the number of detectable line-of-sight perturbers quickly approaches zero, hence our choice of $M_{\rm hm} = 10^9~M_{\odot}$ as an upper limit. 
We chose the prior that, without any knowledge of the Likelihood, is least informative on the scale of $M_{\rm hm}$, that is uniform in $\log$ space. We discuss this choice in Appendix \ref{sec:prior_selection}.

To further increase computing efficiency, we set the quasars to be point sources. For each point on the grid, we calculate the likelihood with a Monte Carlo-based approach as follows:
\begin{enumerate}
\item first, on each grid point of mass function parameters $\vec \theta$, we draw a random set of macro-model parameters $\vec \theta_M$ from $Q(\vec{ \theta}_{\rm M})$ (importance sampling); 
\item then, we draw the corresponding set of line-of-sight haloes and subhaloes from $P(N| \vec{ \theta},\vec{ \theta}_{\rm M})$, we stress here that only the number of substructures depends on $f_{\rm sub}$ via the projected mass of the main lens within an aperture of $2\theta_E$; 
\item for each subhalo and line-of-sight halo we draw its mass, redshift and projected position from $P(\vec{ \theta}_{\rm m}| \vec{ \theta},\vec{ \theta}_{\rm M})$; 
\item we use {\sc pylens}, a {\sc Python}-based ray-tracing package that implements multi-plane lensing with analytical mass profiles, to derive the predicted image fluxes and positions, and calculate the relative likelihood. 
\end{enumerate}
In total, we generate 100\,000 Monte Carlo realizations on each grid point of $\vec{ \theta}$, and the posterior probability is then constructed from the summed likelihood. Since each lens is considered independent, we multiply the likelihood of each lens to obtain a joint inference on the model parameters. A schematic view of this strategy is provided in Figure \ref{fig:flow}.


\begin{table*}
\caption{Summary of the multiply imaged lensed quasars with radio or MIR flux measurements. The references for the lensed image positions and fluxes, and references for the evidence of more complex lens models (e.g., luminous satellites and stellar discs) are also listed. Additional information can also be found on the CASTLES lens database at https://www.cfa.harvard.edu/castles/. The symbol (\dag) indicates those systems that have radio flux-ratio measurements obtained from monitoring \citep{K03}.}
\centering
\begin{tabular}{lccccccc}
\hline
Lens & Type & Radio flux & MIR flux & Satellite & Disc  & References \\
\hline
CLASS~B0128+437{~\dag} & fold & $\surd$ & -- & -- &  -- & \citet{Phillips2000}\\
MG~J0414+0534 & fold & $\surd$ & $\surd$ & $\surd$ & -- & \citet{Falco1997,Minezaki2009}\\
HE\,0435$-$1223  & cross & $\surd$ & --  & $\surd$ & -- & \citet{Wisotzki2002,Jackson2015}\\
CLASS~B0712+472{~\dag} & fold & $\surd$ & -- & -- & $\surd$  & \citet{Jackson1998,Hsueh2017}\\
HS\,0810+2554 & fold & $\surd$ & --  & -- & -- & \citet{Reimers2002,Jackson2015}\\
RX~J0911+0551 & cusp & $\surd$ & --  & $\surd$ & -- & \citet{Bade1997,Jackson2015}\\
PG\,1115+080 & fold& -- & $\surd$ & -- & -- &  \citet{Weymann1980,Chiba2005}\\
CLASS~B1359+154{~\dag } & fold & $\surd$ & -- & -- & --  & \citet{Myers1999,Rusin2001}\\
JVAS~B1422+231{~\dag}& cusp& $\surd$ & $\surd$ & -- & --  & \citet{Patnaik1999}\\
CLASS~B1555+375{~\dag} & fold & $\surd$ & -- & -- & $\surd$  & \citet{Marlow1999,Hsueh2016}\\
CLASS~B1608+656 & fold & $\surd$ & -- & --  & --  & \citet{Koopmans1999,Fassnacht2002}\\
CLASS~B1933+503 & fold & $\surd$ & -- & -- & $\surd$  & \citet{Sykes1998,Suyu2012}\\
CLASS~B2045+265{~\dag} & cusp& $\surd$ & -- & -- & --  & \citet{Fassnacht1999,mckean07}\\
Q2237+030 & cross& -- & $\surd$ & -- & --  &  \citet{Huchra1985,Minezaki2009}\\
\hline
\label{tab:lens}
\end{tabular}
\end{table*}

\begin{table*}
\caption{Summary of the observational data for the seven gravitationally  lensed quasars we used in our analysis. The lensed image positions are in units of arcsec. The values in the parentheses are the corresponding uncertainties. Note that MG~J0414+0534 and PG\,1115+080 do not have flux monitoring data from \citet{K03} and their flux-ratio uncertainties are conservatively assigned to be 10~per cent.}
    \centering
    \begin{tabular}{ll|cc|cc}
    \hline
    Lens & Image & \multicolumn{2}{c}{Positions} & Flux ratio & Reference\\
     & & RA & DEC & & \\
    \hline
     CLASS~B0128+437    & A & $\equiv0$ & $\equiv0$ & $\equiv1$\\
      & B & +0.098 (0.003) &	+0.094 (0.003) & 0.584 (0.029) & \citet{Phillips2000}\\
      & C & +0.520 (0.003) &	$-$0.172 (0.003) & 0.520 (0.029)&\\
      & D & +0.108 (0.003) &	$-$0.250 (0.003) & 0.506 (0.032)&\\
     \hline
     MG~J0414+0534    & A1 & +0.5876 (0.0003) & $-$1.9341 (0.0003) & $\equiv1$ & \\
      & A2 & +0.7208 (0.0003) &	$-$1.5298 (0.0003) & 0.9027 (0.0903)& \citet{Katz1997}\\
      & B & $\equiv0$ & $\equiv0$ & 0.3890 (0.0389)&\\
      & C & $-$1.3608 (0.0007) & $-$1.6348 (0.0008) & 0.1446 (0.0145)&\\
     \hline
     CLASS~B0712+472    & A & $\equiv0$ & $\equiv0$ & $\equiv1$ &  \\
      & B & +0.056 (0.003) & $-$0.156 (0.003) & 0.843 (0.061)& \citet{Hsueh2017}\\
      & C & +0.812 (0.003) & $-$0.663 (0.003) & 0.418 (0.037)&\\
      & D & +1.174 (0.003) & +0.459 (0.003) & 0.082 (0.035)&\\
     \hline
     PG\,1115+080    & A1 & +1.328 (0.003) & $-$2.034 (0.003)& $\equiv1$ & \\
      & A2 & +1.477 (0.004) & $-$1.576 (0.004) & 0.93 (0.093) & CASTLES, \\
      & B & $-$0.341 (0.003) & $-$1.961 (0.003) & 0.16 (0.016)& \citet{Chiba2005}\\
      & C & $\equiv0$ & $\equiv0$ & 0.21 (0.021) &\\
     \hline
     JVAS~B1422+231    & A & +0.38925 (0.00005) &	+0.31998 (0.00005)& $\equiv1$  & \\
      & B & $\equiv0$ & $\equiv0$ & 1.062 (0.009)& \citet{Patnaik1999}\\
      & C & $-$0.33388 (0.00005) &	$-$0.74771 (0.00005) & 0.551 (0.007)&\\
      & D & +0.95065 (0.00005) &	$-$0.80215 (0.00005) & 0.024 (0.006)&\\
     \hline
     CLASS~B1555+375    & A & $\equiv0$ &$\equiv0$ & $\equiv1$ & \\
      & B & $-$0.0726 (0.001) & +0.0480  (0.001) & 0.620 (0.039)& \citet{Hsueh2016}\\
      & C & $-$0.4117 (0.001) & $-$0.0280 (0.001) & 0.507 (0.030)&\\
      & D & $-$0.1619 (0.003) & $-$0.3680  (0.003) & 0.086 (0.024)&\\
     \hline
     CLASS~B2045+265    & A & $\equiv0$ & $\equiv0$ & $\equiv1$ & \\
      & B & $-$0.1338 (0.0001) & $-$0.2483  (0.0001) & 0.578 (0.059)& \citet{mckean07}\\
      & C & $-$0.2877 (0.0001) &	$-$0.7904 (0.0001) & 0.739	 (0.073)&\\
      & D & +1.6268 (0.0002) &	$-$1.0064 (0.0002) & 0.102	(0.025)&\\
    \hline
    \end{tabular}
    
    \label{tab:lens_data}
\end{table*}

\section{The data}
\label{sec:data}

We have collected all radio or MIR observations that are available for the fourteen multiply-imaged quasars that have four lensed images (see Table \ref{tab:lens} for a summary of their general properties). Out of these fourteen lens systems, only seven are used in our full analysis, while the remaining seven are excluded for the following reasons.

\begin{enumerate}
\item HE\,0435$-$1223, HS\,0810+2554, and RX\,J0911+080 are bright optical quasars where faint radio emission was detected by \citet{Jackson2015} from deep Very Large Array (VLA) and e-Multi-Element Remotely Linked Interferometer Network (e-MERLIN) imaging at cm-wavelengths. Although they demonstrate the feasibility of detecting radio emission from radio-quiet quasars, the lensed images are partially resolved, and are, therefore, not suitable to be modelled as a point source or with a single Gaussian component. Very Long Baseline Interferometry (VLBI) observations of HS\,0810+2554 at mas-scale angular resolution has confirmed that the radio structure is indeed extended and composed of multiple compact components \citep{Hartley2019}.
\item CLASS~B1359+154 and CLASS~B1608+656 have multiple lensing galaxies within the Einstein radius and show strong flux-ratio anomalies \citep{Rusin2001,Koopmans1999,Suyu2009}. Moreover, CLASS~B1608+656 is a merging system, which may cause significant bias on the abundance of small-scale structures. Considering the strong coupling between the flux-ratio anomalies generated by multipole components in the macro model and by substructure/line-of-sight haloes, we exclude these two systems from our analysis.
\item  CLASS~B1933+503 is a 10-image system with a face-on spiral as the main lensing galaxy \citep{Sykes1998,Suyu2012}. We notice that the magnification of the lensed images close to the spiral arms has significantly strong deviations from the smooth model predictions. These strong anomalies are very likely due to the presence of the spiral arms. While we exclude this system from our current analysis, we plan to develop an algorithm that includes more complex baryonic structures from simulated disc galaxies into the lens modelling in the future.
\item Q2237+030 is gravitationally lensed by the bulge of a low-redshift spiral galaxy \citep{Irwin1989}. The mass distribution of this system is, therefore, dominated by baryonic structures rather than a smooth dark matter distribution. Hence, as for CLASS~B1933+503, we decided to exclude this system from our current analysis until we develop an appropriate description.
\end{enumerate}

Our final sample includes the following lens systems: CLASS~B0128+437, MG~J0414+0534, CLASS~B0712+472, PG\,1115+080, JVAS~B1422+231, CLASS~B1555+375, and CLASS~B2045+265. Table \ref{tab:lens_data} summarizes the observational data we used in this work (positions and flux-ratios). Improved flux-ratio measurements are available from the MERLIN key programme \citep{K03} for CLASS~B0128+437, CLASS~B0712+472, JVAS~B1422+231, CLASS~B1555+375 and CLASS~B2045+265. These measurements result in an improved flux-ratio uncertainty of less than 5 per cent. For the remaining systems, MG~J0414+0534 and PG\,1115+080,  we adopt a flux uncertainty of 10 per cent. Each lens is modelled with a singular isothermal ellipsoid plus external shear, except for:
\begin{enumerate}
\item MG~J0414+0534 has a luminous satellite (object X) that is detected in optical imaging \citep{Falco1997}, which we include into the lens model and allow its mass to be a free parameter;
\item CLASS~B0712+472 and CLASS~B1555+375 both have an edge-on disc that lies across the merging images, where the flux-ratio anomalies are most significant \citep{Jackson1998,Hsueh2016,Hsueh2017}. We apply the best-fit models found by \citet{Hsueh2017,Hsueh2016} and let the disc mass be the only free parameter. It should also be noted that for the lens system CLASS~B1555+375, the redshift of the lensing galaxy is unknown. A recent detection of an emission line in the NIR spectrum suggests that the source redshift is $z_s = 1.432$ (Fassnacht et al., in prep). Considering the red colour of the lensing galaxy, it is likely to be at high redshift and we assume the lens redshift to be $z_l = 1.0$.  The disc mass fraction of CLASS~B0712+472 and CLASS~B1555+375 is about 15 per cent within the Einstein radius, which is consistent with the range of disc mass fraction in the SWELLS survey \citep{Brewer2014}.
\end{enumerate}

The lens system CLASS~B2045+265 shows a strong de-magnification on the central image of the cusp triplet \citep{Fassnacht1999}. This strong flux-ratio anomaly was thought to be due to the presence of a luminous satellite detected in NIR imaging, although the lens model was peculiar in that the satellite needed to be highly elognated \citep{mckean07}. However, additional Keck adaptive optics imaging at a different epoch has shown proper motion of the luminous object, which indicates that it is very likely to be a star. Therefore, we do not include this additional component in the lens model.

For each lens system, Table \ref{tab:zn} summarizes the redshift of the lens and source (where available), the Einstein radius, and the expected number of substructure and line-of-sight haloes for the idealized CDM model and for a 8.0 keV thermal relic model. As expected from Sections \ref{sec:Substructure} and \ref{sec:LOS}, the mean expected number of subhaloes is determined by $f_{\rm sub}$, $M_{\rm hm}$ and the mass of the host galaxy, while the mean expected number of line-of-sight haloes is a function of $M_{\rm hm}$ and the volume of the light cone, that is, the redshift of the source and the lens. We notice that the expected number of line-of-sight haloes can be significantly larger  than the number of substructures, depending on the length of the light-path. In the next section, we present how the contribution from these additional haloes affects our inference on $f_{\rm sub}$. 

\begin{table*}
\caption{Each column represents (1) the lens name, (2) the lens redshift, (3) the source redshift, (4) the Einstein radius (in arcsec), (5) the opening angle, (6) to (8) the expectation value of substructures in the main halo for the idealized CDM model with $f_{\rm sub} = 0.5$~per cent, 1~per cent, 2~per cent, respectively, (9) the expectation value of line-of-sight haloes for the idealized CDM model within $2\theta_E$, (10) to (12) the expectation value of substructures for a thermal relic model with a particle mass of $8.0$~keV with $f_{\rm sub} = 0.5$~per cent, 1~per cent, 2~per cent, respectively, and (12) the expectation value of line-of-sight haloes for the same thermal relic model. Redshift references: CLASS~B0128+437; \citet{Mckean2004}, MG~J0414+0534; \citet{Tonry1999},  CLASS~B0712+472; \citet{Fassnacht1998},  PG\,1115+080; \citet{Weymann1980,Tonry1998}, JVAS~B1422+231; \citet{Patnaik1999,Impey1996}, CLASS~B1555+375; Fassnacht et al. in prep.; CLASS~B2045+265; \citet{Fassnacht1999,mckean07}.}
\centering
\begin{tabular}{lcccc|cccc|cccc}
\hline
 &  &  &  & & \multicolumn{3}{c}{CDM}&  & \multicolumn{4}{c}{Thermal relic (8.0 keV)} \\
 &  & & & & \multicolumn{3}{c}{$\mu_{\rm sub}$} &  $\mu_{\rm los}$ & \multicolumn{3}{c}{$\mu_{\rm sub}$} & $\mu_{\rm los}$\\
\hline
Lens & $z_l$ & $z_s$ & $\theta_E$ (arcsec) & $\Delta \phi$ (deg.) & 0.5\% & 1\%  & 2\%  & & 0.5\%  & 1\%  & 2\%  & \\
\hline
CLASS~B0128+437 & 1.145 & 3.12 & 0.21 & 123.3 & 7 & 15 & 29 & 240 & 2 & 4 & 6 & 29 \\
MG~J0414+0534 & 0.96 & 2.64 & 1.12 & 101.5 & 631 & 1262 & 2474 & 4666 & 128 & 255 & 510 & 552\\
CLASS~B0712+472 & 0.41& 1.34 & 0.69 & 76.9 & 104 & 209 & 417 & 206 & 21 & 42 & 84 & 25 \\
PG\,1115+080 & 0.3098 &	1.722 & 1.13 & 121.2 & 226 & 452 & 903 & 507 & 45 & 92 & 183 & 61 \\
JVAS~B1422+231 & 0.34 & 3.62 & 0.75 & 77.0 & 99 & 197 & 394 & 395 & 20 & 40 & 80 & 47 \\
CLASS~B1555+375 & (1.0) & 1.432 & 0.25 & 102.6 & 39 & 78 & 157 & 89 & 8 & 16 & 32 &  11 \\
CLASS~B2045+265 & 0.87 &	2.35 & 1.08 & 34.9 & 556 & 1113 & 2225 & 3497 & 112 & 224 & 450 & 414 \\
\hline
\end{tabular}
\end{table*}\label{tab:zn}


\section{results}
\label{sec:results}

 We present the results of our analysis in two parts. In Section \ref{sec:fsub_res}, we present our constraints on the mass fraction under the assumption of an idealised CDM model (i.e. $M_{\rm hm} = 0$) and how the constraints change with the inclusion of line-of-sight haloes. In Section \ref{sec:mhm_res}, we focus on thermal relic dark matter models, which also include the WIMP CDM model at $M_{\rm hm} = 10^{-6}~M_\odot$. In Section \ref{sec:sys}, we discuss the effect of systematic uncertainties on our results.

\begin{figure*}
\centering
\includegraphics[scale=0.5]{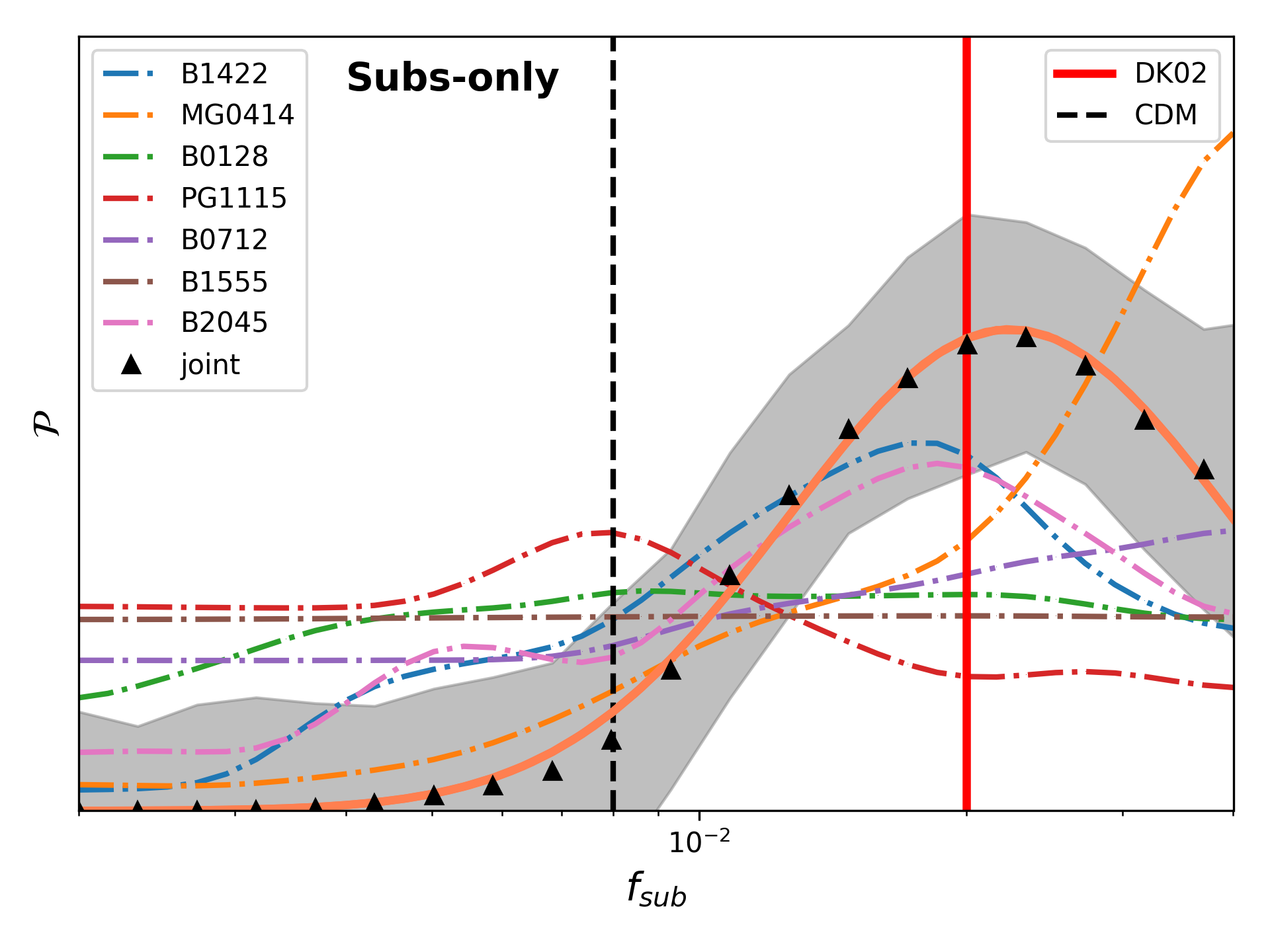} 
\includegraphics[scale=0.5]{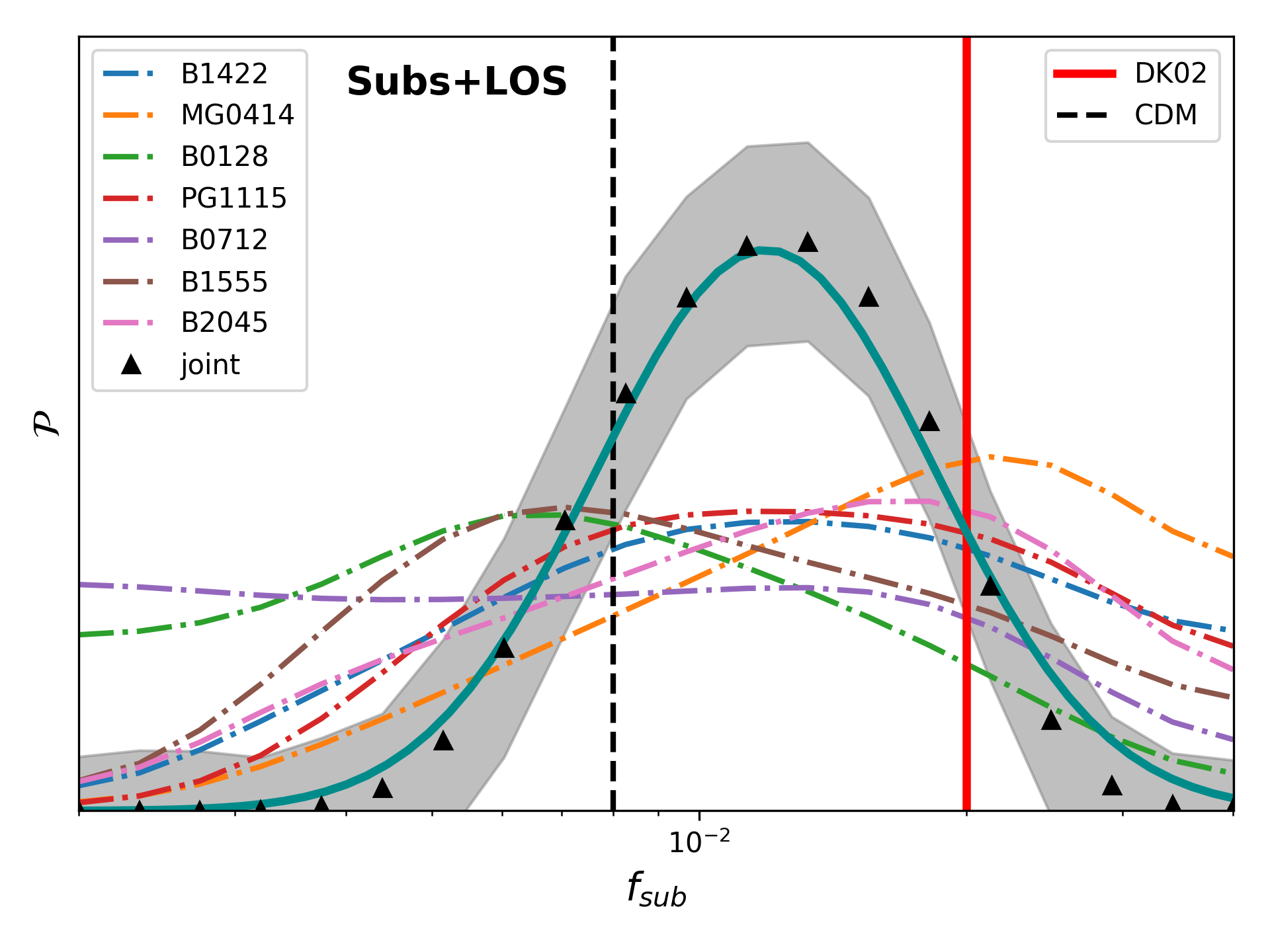} 
\caption{Joint constraints (triangles and thick curve) on the substructure mass fraction ($f_{\rm sub}$) without (left) and including (right) line-of-sight haloes from seven gravitationally lensed quasars, under the assumption of an idealized CDM model. The shaded area shows the 1-$\sigma$ uncertainties of the probability distribution. The red solid vertical lines show the median constraint results from \citet{Dalal2002}, and the black dashed vertical lines show the upper-limit from numerical simulations \citep{Xu15}.}\label{fig:fsub}
\end{figure*}

\begin{figure}
\centering
\includegraphics[scale=0.5]{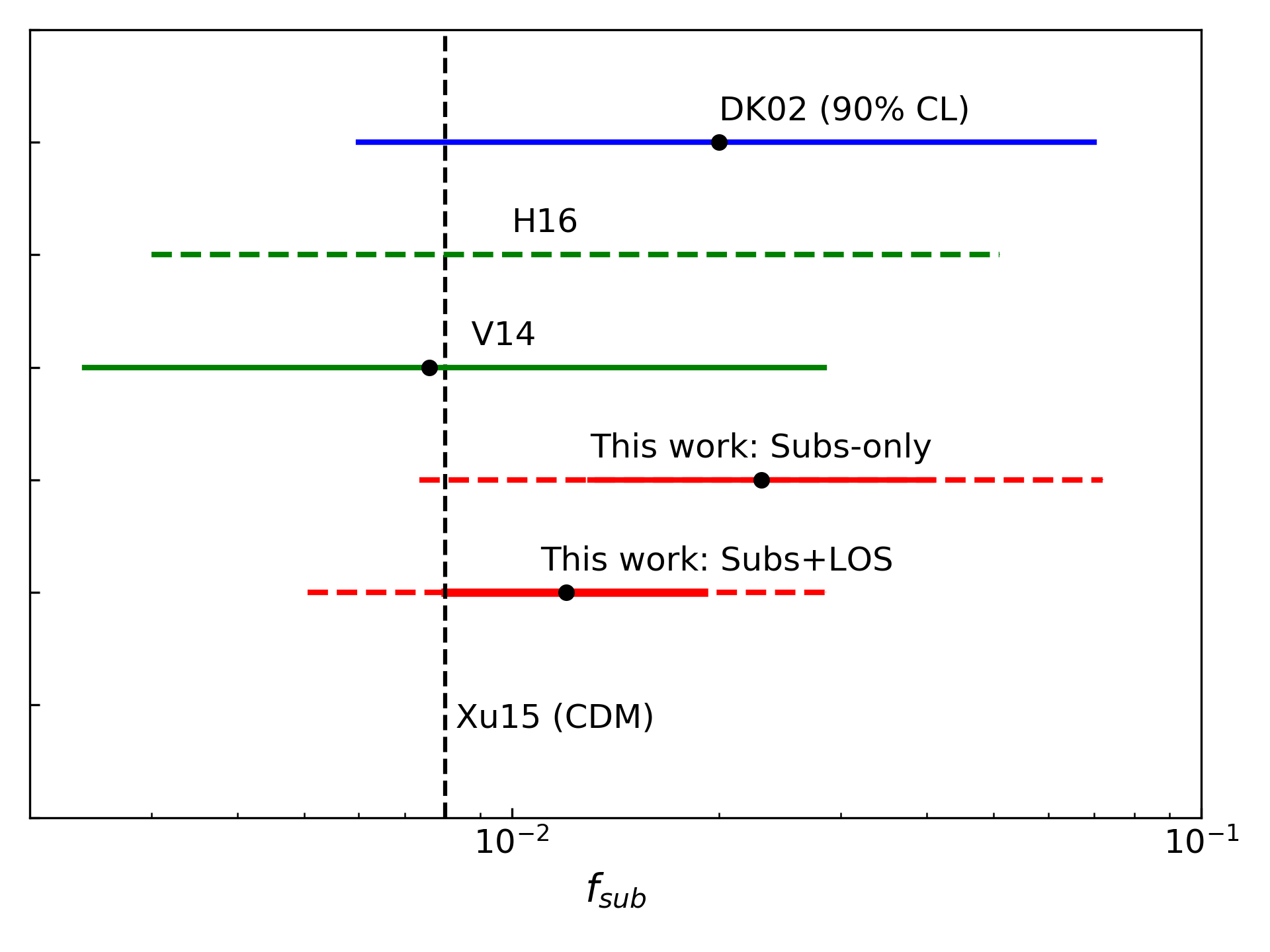}
   \caption{Constraints on $f_{\rm sub}$ (CDM-only) derived in this paper and other strong lensing studies. DK02: \citet{Dalal2002} analysis of seven lensed quasars; H16: \citet{Hezaveh16} analysis of SDP.81 ALMA observations using the gravitational imaging technique; V14: \citet{V14a} analysis of 11 SLACS lenses using the gravitational imaging technique;  Subs+LOS \& Subs-only: results from this work {\it with} and {\it without} line-of-sight haloes, respectively. Xu15: \citet{Xu15} analysis of re-scaled haloes in a CDM-only $N$-body simulation (Aquarius). All uncertainties are presented at the 68~per cent CL (solid) and 95~per cent CL (dashed), except for DK02, which is at the 90~per cent CL).}
\label{fig:summary_fig1}
\end{figure}

\subsection{Substructure mass fraction (CDM only)}
\label{sec:fsub_res}

In this section, we present the results of our analysis  {\it with} and {\it without} line-of-sight haloes under the assumption of an idealized CDM cosmology. Our substructure-only analysis demonstrates the improvements on the precision of the constraints due to the more accurate flux-ratio measurements and lens macro models that are now available. Importantly, we find that the inclusion of line-of-sights haloes now resolves the tension between the prediction from numerical simulations 
\citep{Mao2004,Xu15} and the large value of $f_{\rm sub}$ inferred by \citep{Dalal2002}.

\subsubsection{CDM substructure-only}
\label{sec:fsub_cdm}

We first derive constraints on the substructure mass fraction assuming a CDM model, that is, $M_{\rm hm}=0$, and excluding the contribution from line-of-sight haloes.
The corresponding posterior probability distribution is presented as the curve marked with triangles in Figure~\ref{fig:fsub}, where the error bars represent the 1-$\sigma$ uncertainties on the Monte Carlo integrals of the probability in each bin. From a Gaussian fit to the posterior curve of each lens, we derive a mean joint value of $f_{\rm sub} = 0.023^{+0.018}_{-0.010}$ at the 68 per cent confidence level (CL). We find this value to be larger than, but within 2-$\sigma$ of, what is predicted by CDM-only numerical simulations for host galaxies with similar masses and redshifts, and substructure masses in the same range \citep[$f_{\rm sub} = 0.008$;][the value is recalculated to fit the definition of $f_{\rm sub}$ in this work]{Xu15}. We stress that due to galaxy formation processes, where baryonic effects may suppress the level of substructure, this discrepancy may be larger \citep{Despali2017}. 

Interestingly, from a sample of eleven galaxies from SLACS, \citet{V14a} have inferred a fraction of $f_{\rm sub} = 0.0064^{+0.0080}_{-0.0042}$, which is slightly smaller than, but is in much closer agreement with numerical predictions from both dark-matter-only and hydro-dynamical CDM simulations \citep{Despali2017}. Although \citet{V14a} infer a much smaller $f_{\rm sub}$ than our substructure-only result, we emphasize that the SLACS lenses are at relatively low redshift, and therefore, are less affected by line-of-sight haloes than our higher-redshift sample of lensed quasars. 

Our results are also consistent with those of \citet{Dalal2002}, who analyzed a sample of lenses of comparable size and with some overlap in terms of lens systems. The improved precision of our results and an upper limit that is much closer to the theoretical predictions are mainly due to an improvement in the flux-ratio uncertainties from 20 to around 5 per cent. The improvement clearly demonstrates the importance of acquiring  accurate flux measurements for more lensed quasars in the future. We also emphasize that, unlike \citet{Dalal2002}, we include in the macro-models of CLASS B0712+472 and CLASS B1555+375 an edge-on stellar disc that can explain most of the observed flux-ratio anomalies. The inclusion of this extra component brings down the upper limit on $f_{\rm sub}$ that had been inferred when fitting a single SIE model to these data.

From Figure~\ref{fig:fsub}, we notice that MG~J0414+0534 is essentially an outlier, with a posterior probability that peaks at large values of $f_{\rm sub}$, outside our prior range. It is one of the most observed lensed quasars with a wide frequency coverage from radio, sub-millimetre molecular lines, far-infrared to the MIR. Recently, \citet{Stacay2018} have shown that the flux ratios for this system do not change with frequency. As different wavelengths should be sensitive to the different mass scale of the perturbation (due to a change in the source size), we conclude that the large anomaly observed in MG~J0414+0534 is more likely due to more massive structures, for example, a more complex macro model. Further investigation and potentially deeper data for this system are required to conclusively understand the origin of the anomaly. When we exclude this system, we infer a mean mass fraction of $f_{\rm sub} = 0.019^{+0.008}_{-0.009}$ at the 68 per cent CL.

As discussed in Section \ref{sec:data}, we have not included the presence of a previously thought luminous satellite in the lens system CLASS~B2045+265, given that it has now been identified as a foreground star from proper motion. If, together with MG~J0414+0534, we also exclude this system, then we obtain $f_{\rm sub}=0.018^{+0.013}_{-0.008}$, in agreement within 1-$\sigma$ of the expectations from CDM-only numerical simulations. We conclude, therefore, that complex macro-models can have a significant impact on the correct interpretation of flux-ratio anomalies, and that further investigations of the lens systems MG~J0414+0534 and CLASS~B2045+265 are required. A summary of the results from the different choices of sample sets is given in Table \ref{tab:con_summary}.

\subsubsection{CDM substructure and line-of-sight haloes}

\citet{Metcalf05} and \citet{Gilman2019} have shown that low-mass haloes located along the lines of sight of the lens galaxies can have a dominant effect on the relative fluxes of multiply-imaged quasars \citep[see][for a similar result for extended lensed images]{Despali2018}. In this section, we discuss how our inference on the substructure mass fraction changes when the contribution from this population is taken into account, which is also shown in Figure~\ref{fig:fsub}. 

As expected, we find a significant drop of 50 per cent in the substructure mass fraction, to $f_{\rm sub} = 0.012^{+0.007}_{-0.004}$, at the 68 per cent CL. The result is a reflection of the fact that once the dominant contribution from the line-of-sight haloes is included, a smaller abundance of substructure is needed to reproduce the observed flux ratios. In particular, our results imply that about half of the flux-ratio anomalies are produced by line-of-sight structures. This result may also explain the findings of \citet{V14a}: if the degeneracy with the line-of-sight halo population has a significant impact on the substructure mass fraction inference, this effect is expected to be larger for the sample of high redshift lensed quasars considered here, than for the SLACS sample, because of the larger cosmological volume probed by the former.

Our new constraints on the total mass fraction in substructure are now consistent with CDM-only numerical predictions at the 1-$\sigma$ level \citep[$f_{\rm sub}$ = 0.008;][]{Xu15}. \citet{Despali2017} have quantified the suppression in the number density of substructure in hydro-dynamical simulations  relative to N-body-only simulations to be between 20 and 40 per cent, and the drop in the substructure mass fraction to be between 40 and 70 per cent. According to this correction, our constraints are also in agreement with CDM-hydrodynamical simulations at the 1.2-$\sigma$ level. When we exclude MG~J0414+0534, we infer $f_{\rm sub} = 0.010^{+0.005}_{-0.004}$, and when we exclude both MG~J0414+0534 and CLASS~B2045+265, we infer $f_{\rm sub} = 0.009^{+0.006}_{-0.004}$. Table \ref{tab:zn} also shows that the expectation value of substructures and line-of-sight haloes are at the same level when $f_{\rm sub} = 0.01$, for a CDM cosmology.

Figure~\ref{fig:summary_fig1} presents the comparison between the results from this paper and other studies that constrain $f_{\rm sub}$ from gravitational lensing. In this work, we define $f_{\rm sub}$ to be the substructure fraction within the aperture of $2\theta_E$. In \citet{Dalal2002}, $f_{\rm sub}$ is defined as one half of the convergence at the critical radius, which is also the same definition used by \citet{V14a} and \citet{Hezaveh16}. This definition requires the perturbers to be close to the critical radius of the lens, which holds for the simulation designs of \citet{Dalal2002} and the gravitational imaging technique. Following \citet{Xu15}, we recalculate the \citet{Dalal2002} and \citet{V14a} results to match our definition of $f_{\rm sub}$. The main difference between the definitions is that our work probes the gravitational effects of perturbers within a larger aperture, that is, to $2\theta_E$. However, these numbers are considered comparable within the region the convergence is close to the critical radius, as the projected positions of subhaloes are uniformly distributed. We also emphasize that, although the current estimates on $f_{\rm sub}$ scatter from $10^{-3}$ to $10^{-2}$, all of the estimates are marginally in agreement within the 1- to 2-$\sigma$ level.
 
\subsection{Inference on thermal relic dark matter}
\label{sec:mhm_res}

\citet{Sch2012} and \citet{Lovell2014} have shown that the effect of the free streaming of thermal relic dark matter particles can be well described by the half-mode mass $M_{\rm hm}$. In terms of the subhalo mass function, this introduces an extra free parameter that is degenerate with the substructure mass fraction. Suppression in the number of low-mass haloes can either be related to a larger value of the half-mode mass or conversely to a lower dark matter fraction in the substructure. While the value of the half-mode mass solely depends on the dark matter physics, $f_{\rm sub}$ is related to the accretion history of the host halo and the efficiency of tidal disruption. 

As we have assumed a mean line-of-sight halo mass function, we have ignored any degeneracy between the halo mass function normalization and $M_{\rm hm}$. We expect that this degeneracy does not cause a significant effect on the results of this paper, as the systematic uncertainties in the macro-model for those galaxies that lack deep imaging information should be larger than the scatter introduced by varying lines of sight.

Under the assumption of a thermal relic dark matter model we infer a mean dark matter fraction of $f_{\rm sub} = 0.013\pm 0.007$, where the slightly increased value relative to our CDM-only results can be explained by the non-zero value of the half-mode mass. At present, we are not aware of any numerical simulation on cosmological scales of massive galaxies with a thermal relic cosmology that resolves small-scale haloes. Therefore, we cannot compare our $f_{\rm sub}$ results with theoretical expectations. In the following, we present the marginalized constraints on $M_{\rm hm}$ and discuss the implications for the free streaming properties of dark matter. In this part of the analysis, we have included the contribution of both substructure and line-of-sight haloes.

\begin{figure*}
\centering
\includegraphics[scale=0.5]{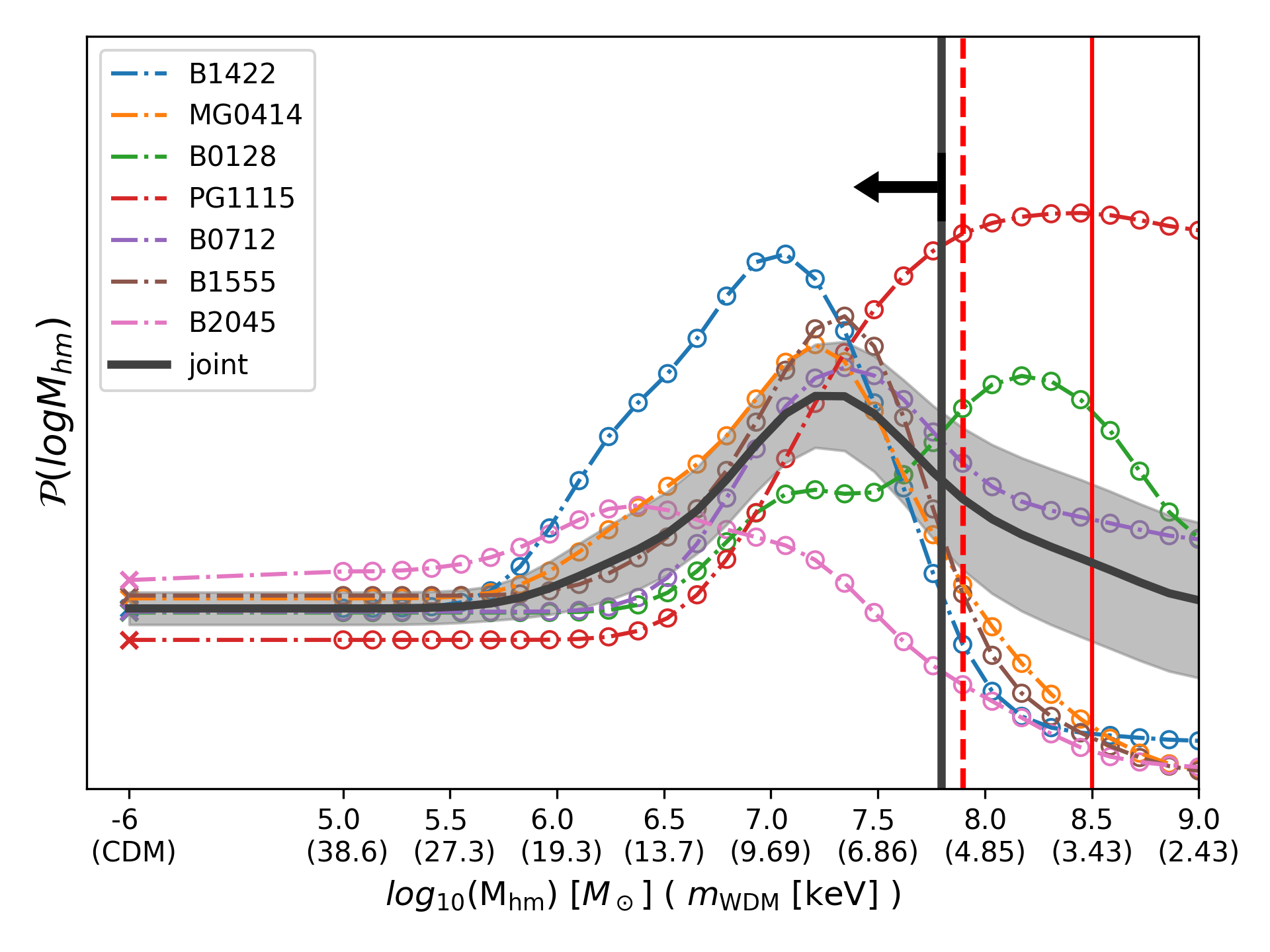}
\includegraphics[scale=0.5]{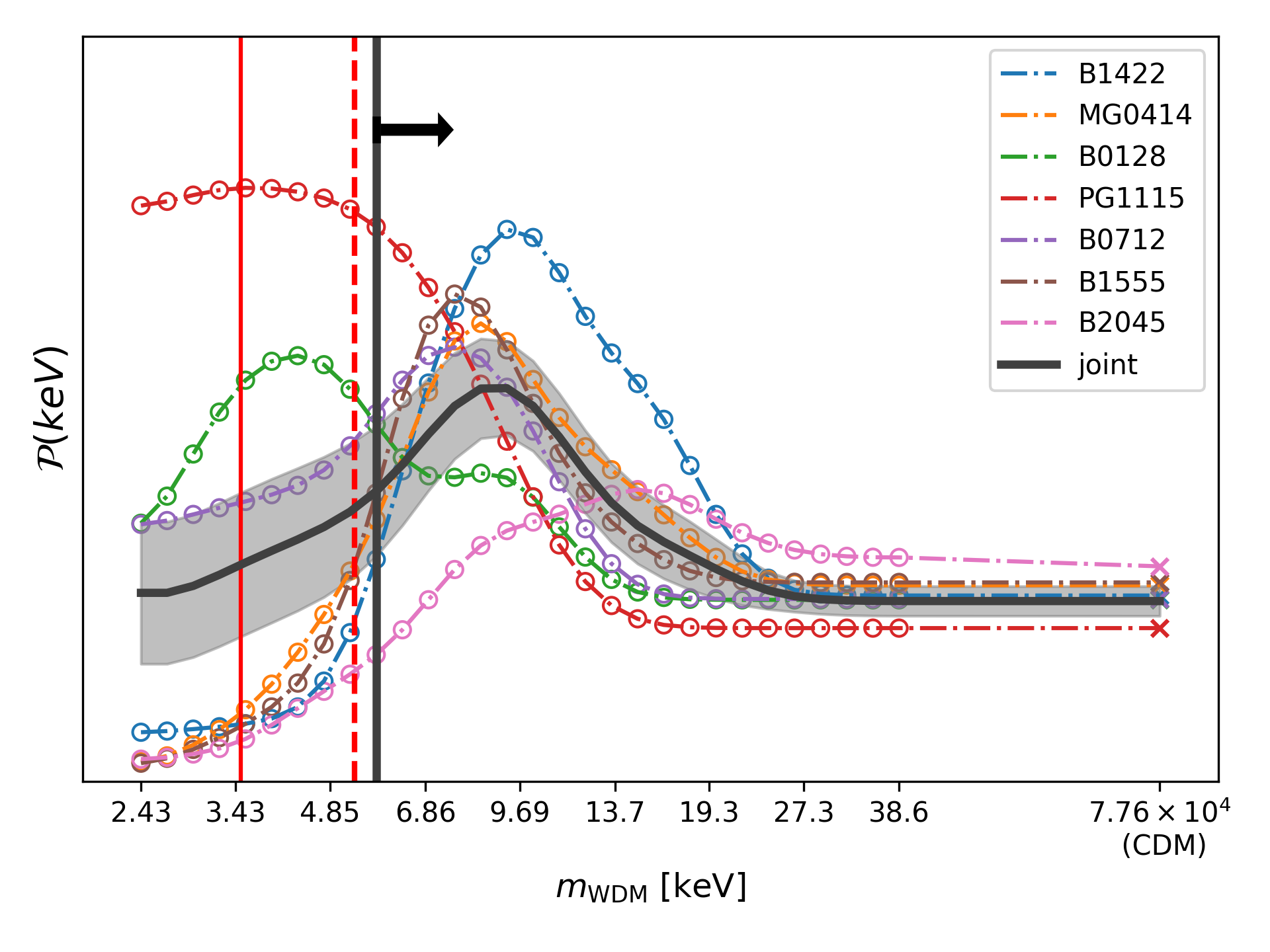}
 \caption{The joint posterior probability distribution for the half-mode mass (left) and thermal relic particle mass (right) from our sample of seven gravitationally lensed quasars. The grey shaded area represents the 1-$\sigma$ uncertainty of the joint constraint. The black vertical line represents our upper limit on $M_{\rm hm}$ and lower limit on $m_{\rm th}$ at the 95~per cent CL. The black arrows show the direction of the allowed region at the 95~per cent CL from this work. The red dashed and solid lines represent the lower limits (95~per cent CL) from the latest Ly$\alpha$ forest constraints, assuming a smooth and non-smooth intergalactic medium temperature evolution, respectively \citep{Irsic2017}.  }
\label{fig:mwdm}
\end{figure*} f

Figure~\ref{fig:mwdm} presents our constraints on the free-streaming property of thermal relic dark matter from our sample of seven gravitationally lensed quasars. In both panels, the grey shaded area shows the $1-\sigma$ uncertainty of the Monte Carlo integral on the probability distribution of the joint constraint.  As anticipated, the likelihood function for all lenses flattens to a constant value for dark matter models colder than about $M_{\rm hm} \leq 10^{5.5} ~M_{\odot}$ due to the small difference in the number of haloes that can be detected between the different dark matter models. Similarly, we expect the likelihood to reach a constant value above $M_{\rm hm} =10^{9} ~M_{\odot}$ as the number of detectable perturbers is equally consistent with zero for all models.

From Figure~\ref{fig:mwdm}, we also notice that the lens systems PG\,1115+080 and CLASS~B0128+437 show a clear preference for warmer models, especially towards the region where no detectable perturber is expected. This is due to the fact that these two systems show no flux-ratio anomaly, so that a smooth macro lens model can successfully predict the image flux ratios. In contrast, the posteriors of other lenses increase toward the colder region until the limit of the data sensitivity. 
We interpret this bi-modal distribution as the result of small-sample statistics. Despite this, our results demonstrate that the analysis of gravitationally lensed quasars is a promising approach to further explore the parameter space of dark matter models.

Our joint analysis results in an upper limit at the 95 per cent CL for the half-mode mass of $M_{hm}< 10^{7.8}$~M$_{\odot}$, which corresponds to a lower limit on the thermal relic mass of $m_{\it th} > 5.6$~keV at the same CL.
In comparison with the latest 2-$\sigma$ constraints from the Ly$\alpha$ forest, $m_{\it th}>5.3$~keV or $m_{\it th}>3.5$~keV, depending on the assumption of the intergalactic medium thermal history \citep{Irsic2017,Garzilli2018,Bolton2008}, our current results  provide a comparable level of constraints on the mass of the thermal relic dark matter particle, with the potential of significantly improving with upcoming large sample of lensed quasars \citep{Gilman2019}.  Our results are also in agreement with a recent analysis of gravitationally lensed quasars by \citet{Gilman2019b}. However, due to a different choice of prior on the half-mode mass, at present, it is not clear how robustly the two results can be compared with each other.

\begin{table*}
    \centering
    \caption{Summary of our constraints on the dark matter mass fraction $f_{\rm sub}$ (idealised CDM-only), the thermal relic particle mass $m_{\rm th}$ and the half-mode mass $M_{\rm hm}$ for different sub-samples of lensed quasars.}
    \begin{tabular}{r|l|l|l|l}
    \hline
    Sample & \multicolumn{2}{c}{$f_{\rm sub}$ (CDM-only)} & $m_{\rm th}$ &$M_{\rm hm}$ \\
    \hline
    7 quasar lenses & $0.012 ^{+0.007}_{-0.004}$ (subs+LOS) & $0.023 ^{+0.018}_{-0.010}$ (subs-only) & $>5.58$ keV (95\% CL.) & $< 10^{7.80}$~M$_{\odot}$ (95\% CL.)\\

        & &\\
    Exclude MG~J0414+0534 & $0.010 ^{+0.006}_{-0.004}$  & $0.019 ^{+0.008}_{-0.009}$  & $>5.30$ keV  & $< 10^{7.89}$~M$_{\odot}$\\
    & &\\
    Exclude MG~J0414+0534 & $0.009 ^{+0.006}_{-0.004}$  & $0.018 ^{+0.013}_{-0.008}$   &$>4.77$ keV  & $< 10^{8.03}$~M$_{\odot}$\\
    {\& CLASS~B2045+265} & & \\
    \hline
    \end{tabular}
    
    \label{tab:con_summary}
\end{table*}

\subsection{Systematic uncertainties}
\label{sec:sys}

Throughout our analysis, there are several factors that can introduce systematic uncertainties into our inference. We discuss these factors in this section and how they can be addressed in the future.

\subsubsection{Stellar structures} 

Here, we have already included edge-on discs as a higher-order component in the lens models for the cases of CLASS~B0712+472 and CLASS~B1555+375, since these two lenses show solid evidence of stellar discs in high-resolution IR imaging \citep{Hsueh2016,Hsueh2017}. However, \citet{Gilman2018} have shown that even early-type galaxies can have disc-like structures that can contribute to flux-ratio anomaly signals. An analysis of a sample of lens-like galaxies from the Illustris simulation also suggests that stellar structures in early-type lenses can increase the level of flux-ratio anomalies by around 10~per cent \citep{Hsueh2018}. In our sample, Keck adaptive-optics IR imaging of  CLASS B0128+437 shows evidence of a face-on late-type galaxy \citep{Mckean2004,Lagattuta2010} and the lens model for CLASS~B2045+265 requires a significant elliptical component to the mass model \citep{mckean07}. By not accounting for possible stellar structures in the lens modelling, our analysis could potentially over-estimate the abundance of low-mass haloes, and hence artificially favour colder dark matter models. Although evaluating the impact from undetected stellar structures is not possible with current observations, combining the stellar structures from the latest hydro-dynamical simulations of lens-like galaxies, as for example, with the Illustris TNG suite of simulations, into the analysis or obtaining kinematic information on the lensing galaxies will help to reduce this source of systematic uncertainty. This will be the focus of a future theoretical work.
    
\subsubsection{Source structures}

To optimize the computing efficiency, we have assumed the source quasars to be with point-like. There is evidence from VLBI observations that CLASS B0128+437 \citep{biggs04}, MG J0414+0534 \citep{ros00} and CLASS B1555+375 \citep{Hsueh2016} have extended background sources on mas-scales, but CLASS B0712+472 \citep{Hsueh2017}, JVAS~B1422+231 \citep{Patnaik1999} and CLASS~B2045+265 \citep{mckean07} have compact structures; VLBI data for PG\,1115+080 has been taken, but is yet unpublished.  The size of the source affects the sensitivity of the data to small-scale structures, with larger sources being less sensitive to low-mass haloes. Whether this effect results in an artificial and significant preference for warmer dark matter models (potentially compensating for the effect of stellar structures) is unclear, as the size and internal structures of the quasar source can vary from one lens system to the other. In the near future, multi-wavelength observations at radio and sub-millimetre wavelengths will help us to gain further information on the size and structure of the sources and correctly include them in our analysis.

\subsubsection{Source variability} 
\label{sssec:source}

Any variation of the background radio source flux will be seen at different times in the multiple lensed images, meaning that flux measurements taken at a single epoch are sampling the intrinsic light curve of the quasar at different times for the different images.  For this reason, in previous studies, such as \citet{Dalal2002}, the flux-ratio uncertainties were assigned to be 20~per cent for all of the systems with one-time flux measurements. The systematic uncertainties from quasar variability can be eliminated by averaging over a long period of monitoring. In this work, we quote the average flux-ratios from \citet{K03}, which bring the uncertainties in the intrinsic variation from 20~per cent to less than 5~per cent. Most of the radio-sources in our sample have also shown little variation throughout the monitoring.  The clear improvement in the constraints produced by the analysis described in Section~\ref{sec:fsub_cdm} emphasizes the importance of monitoring observations or some other technique for improving the precision of the flux-ratio measurements.
    
\subsubsection{Propagation effects} 

Although propagation effects, such as free-free absorption and scatter broadening can alter the properties of the different lensed images measured at radio-wavelengths, these effects have a strong wavelength dependence, and therefore, can be identified and corrected for with multi-wavelength observations (e.g. \citealt{winn04,biggs03,mittal2007}). There is no clear evidence of propagation effects in our sample, except for CLASS B0128+437 \citep{biggs04} where there is scatter broadening of the lensed images on VLBI scales. It is expected that the contribution from propagation effects to the systematic uncertainties in our analysis are small, but recently completed multi-frequency imaging campaigns with the VLA and VLBI will resolve this.

\vfill\null
\subsubsection{Properties of subhaloes and field haloes}

We have assumed both field haloes and subhaloes to be well described by spherical NFW profiles with a concentration-mass relation from \citet{Duffy2008}. While this is a reasonable assumption for small dark isolated field haloes, the profile of a subhalo may be affected by its interaction with the host halo and depends on its distance from the host centre and its properties at infall. In general, subhaloes are expected to be more concentrated than isolated haloes, especially those closer to the host centre due to tidal truncation \citep{hayashi03,Springel2008,moline17}. However, \citet{Despali2018} found that, in terms of the lensing efficiency, neglecting  the  difference  in the concentration-mass relation between subhaloes and field haloes leads to an uncertainty in the inferred mass of the order of 5 to 20 per cent for subhalo masses between $10^{6}$ and $10^{9}M_{\odot}$, and an even smaller effect is caused by neglecting the dependence on distance. For this reason, we believe that neglecting the details of the subhalo concentration introduces only a second-order effect in our analysis. Moreover, recent works have shown that the properties of simulated subhaloes and the details of the disruption process, which influence the density profiles, are still not well constrained in numerical simulation at the low-mass end \citep{vandenBosch18a,vandenBosch19b,vandenBosch19,errani19}.

Another relevant aspect is the effect of warm dark matter on the concentration-mass relation. Previous works \citep{ludlow16} have shown that the concentration of low-mass (sub)haloes is depleted with respect to CDM, in a way that is similar to the effect on the mass function. This might result in a smaller lensing signal by low-mass (sub)haloes in WDM models, which we have not considered here. However, the (sub)halo mass is the main parameter that determines the strength of the lensing signal and the concentration has been shown to have a second order effect \citep{Despali2018} and thus we believe that our results are not significantly affected by not including a more complex description of the concentration-mass relation. Moreover, the concentration-mass relation has so far only been constrained for a limited number of specific WDM realisations, and a detailed parametrisation as a function of the WDM model is still lacking. 

Throughout the analysis, we have fixed the number density of all line-of-sight realisations to the average density of the Universe and thus neglected the fact that some lines of sight might be over- or under-dense. In a more stringent analysis, these realisation should reflect the environment of each lens and the characteristic variance of line-of-sight structures throughout the Universe. However, we have information on the environment and the field galaxies only for a few quasar lenses in our sample. Moreover, the link between the number density of luminous field galaxies and dark field haloes is ambiguous. In the future, we plan to use high-resolution large-scale simulations to quantify any variation between different lines of sights, and we will investigate the assumption that field haloes are isolated and do not contain significant subhaloes.

Finally, we parametrised the WDM mass function following the results of numerical simulations \citep{schneider12,lovell12}. We extrapolated the fitting functions from these works below the resolution limit of the simulations ($M<10^{7}M_{\odot}$), where the functional forms result in a sharp decrease in the number of objects. It remains to be  shown  definitively whether  this  drop-off  rate  describes the WDM models accurately and higher-resolution simulations would be required to confirm it.


\section{Conclusions}
\label{sec:conclusions}

We have analyzed a sample of seven gravitationally lensed quasars and used the observed image positions and relative fluxes to probe the abundance of low-mass haloes within the potential of the lensing galaxies and along their lines of sight. Our results can be summarized as follows.

\begin{enumerate}

\item We find that accurate flux ratio measurements are a key ingredient for the derivation of precise constraints on the (sub)halo mass function. By improving the flux-ratio uncertainties from 20 to better than 5 per cent, we substantially bring down both the upper limit and the uncertainty on the normalization of the subhalo mass function, when compared to a previous study by \citet{Dalal2002}, based on a sample of comparable size.

\item Under the assumption of an idealized CDM model, we find that the degeneracy between the substructure and line-of-sight haloes has a significant effect on the inferred substructure mass fraction. In particular, the inclusion of a line-of-sight population brings our constraints on $f_{\rm sub}$ into much closer agreement with the expectations from both CDM-only and hydro-dynamical simulations. This result also explains the long-standing discrepancy between the dark matter fraction inferred by \citet{Dalal2002} and numerical simulations \citep{Xu15}.

\item The inclusion of extra complexity in the mass model of the lensing galaxies, although sub-dominant for the sample considered here, also plays an important role for a correct interpretation of flux-ratio anomalies. In particular, the inclusion of a stellar disc in the macro model makes the edge-on disc lenses no longer outliers in terms of the strength of the observed anomalies. Deep imaging observations are therefore crucial to break the degeneracy between the stellar structures and small-scale dark matter perturbers.  However, as the effect of other complex structures at a smaller scale, such as spiral arms and the intrinsic un-smoothness of elliptical galaxies, has not been properly quantified yet, we plan to address this issue in a follow-up paper.

\item Under the assumption of a thermal relic dark matter model, we constrain the dark matter particle mass to be  $m_{\rm th}>5.6$~keV at the 95 per cent confidence level. Our limits are in agreement with observations of the Ly$\alpha$ forest \citep{Irsic2017}, showing that the study of gravitationally lensed quasars can provide comparably strong constraints with the current sample size. Furthermore, compared to the uncertainties in the intergalactic medium thermal history, gravitationally lensed quasars provide a more direct and robust constraint on dark matter properties, which will be further improved with the analysis of increasingly larger sample of lens systems.
\end{enumerate}

\section*{Acknowledgements}
We thank  Andrew Robertson,  Alexey Boyarsky, Mark Lovell, Simon Birrer, Daniel Gilman, Chuck Keeton, and Dandan Xu for helpful discussions on this work. We also thank Ethan Nadler for important feedback on our preprint. CDF and JWH acknowledge support from the National Science Foundation under Grant No. AST-1715611. JWH and JPM acknowledge support from the Netherlands Organization for Scientific Research (NWO)  (Project No. 629.001.023), Chinese Academy of Sciences (CAS) (Project No. 114A11KYSB20170054), and the Dutch National Supercomputer Service. SV has received funding from the European Research Council (ERC) under the European Union's Horizon 2020 research and innovation programme (grant agreement No. 758853). LVEK acknowledges support by a VICI grant (Project No. 614.001.206) from NWO.




\bibliographystyle{mnras}
\bibliography{SHARPVII.bbl} 



\appendix

\section{Prior selection}
\label{sec:prior_selection}

Choosing a suitable prior is important in Bayesian inference.
In this work we chose a prior that is uniform in $\log (f_{\rm sub})$, because it is shows the least information about the scale of a quantity. But in principle we could have chosen alternative forms for the prior.

A popular choice for priors is given by the Jeffreys prior \citep{jeffreys46}. Calculating the Jeffreys prior for the full Likelihood $\mathcal{P} (\vec d |M_{\rm hm},f_{\rm sub} )$ requires a marginalization over data realizations $\vec d$, which implies a large number of model evaluations.
As this is computationally prohibitive one might only consider the substructure part of the likelihood $\mathcal{P}( \vec \theta_m , N | M_{\rm hm},f_{\rm sub})$.
%
%
The Jeffreys prior in this case has the form:
\begin{equation}
\mathcal{P}(M_{\rm hm},f_{\rm sub}) \propto M_{\rm hm}^{-1} f_{\rm sub}^{-1} ~ \mu(M_{\rm hm},f_{\rm sub}) ~\sqrt{ \mu(M_{\rm hm},f_{\rm sub}) + C}\,,
\end{equation}

with $C \approx 2.3658 \times 10^{10}$. Due to the large value of $C$ this can be approximated as:
$\mathcal{P}(M_{\rm hm},f_{\rm sub}) \approx  M_{\rm hm}^{-1} f_{\rm sub}^{-1} ~ \mu(M_{\rm hm},f_{\rm sub}) \propto \frac{1}{M_{hm}^2}$.

The advantage of the Jeffreys prior is that by construction every other parametrization of the mass function that is different to $(M_{\rm hm}, f_{\rm sub})$ will have a Jeffreys prior that is related via a multiplication of the determinant of the Jacobian (while conjugate priors, for example, do not necessarily stay conjugate priors after a coordinate transformation). This property makes Jeffreys priors quite useful for parameters describing scales of some quantity.


\label{lastpage}
\end{document}